\documentclass[conference]{sty/IEEEtran}
\usepackage{url}
\makeatletter
\g@addto@macro{\UrlBreaks}{\UrlOrds}
\makeatother
\def\UrlBreaks{\do\/\do-}
\DeclareUnicodeCharacter{03BB}{\ensuremath{\lambda}}

\usepackage{amsmath}
\usepackage{booktabs}
\usepackage{subcaption}
\usepackage{microtype}
\usepackage{xspace}
\usepackage[table]{xcolor}
\usepackage{graphicx}
\usepackage{titlesec}
\usepackage{adjustbox}
\usepackage[flushleft]{threeparttable}
\usepackage{multirow}
\usepackage{balance}

\usepackage{hyperref}

\newcommand{\sys}{\mbox{\textsc{Reducio}}\xspace}
\newcommand{\syshelper}{\mbox{\textsc{\sys-Coordinator}}\xspace}
\newcommand{\sysmonitor}{\mbox{\textsc{\sys-Monitor}}\xspace}

\usepackage[T1]{fontenc}

\newcommand{\cc}[1]{\mbox{\texttt{#1}}}

\def\Snospace~{\S{}}

\usepackage{courier}
\usepackage[all]{nowidow}

\input{glyphtounicode}
\usepackage{tikz}
\newcommand*\WC[1]{%
  \begin{tikzpicture}[baseline=(C.base)]
    \node[draw,circle,inner sep=0.2pt](C) {#1};
  \end{tikzpicture}}

\usepackage{xstring}
\newcommand{\PP}[1]{
  \vspace*{2.5pt}
  \noindent{\bf \IfEndWith{#1}{.}{#1}{#1.}}
}

\newenvironment{packeditemize}{
  \begin{list}{$\bullet$}{
    \setlength{\itemsep}{2pt}
    \setlength{\labelwidth}{8pt}
    \setlength{\leftmargin}{10pt}
    \setlength{\labelsep}{3pt}
    \setlength{\listparindent}{\parindent}
    \setlength{\parsep}{1.5pt}
    \setlength{\parskip}{1.5pt}
    \setlength{\topsep}{1.5pt}}}{\end{list}}

\usepackage{etoolbox}
\makeatletter
\patchcmd{\ttlh@hang}{\parindent\z@}{\parindent\z@\leavevmode}{}{}
\patchcmd{\ttlh@hang}{\noindent}{}{}{}
\makeatother

\begin{document}
\title{\sys: Optimized Confidential Serverless\\Cloud Deployments for Enterprise Customers}

\author{
Vikram Ramaswamy{$^{\dagger}$}
\quad
Chuqi Zhang{$^*$}
\quad
Adil Ahmad{$^{\dagger}$}
\\
\textit{\textsuperscript{$\dagger$}Arizona State University
\quad
\textsuperscript{$*$}National University of Singapore
\quad
}
}
\date{}
\maketitle
\thispagestyle{empty}
\sloppy

\begin{abstract}

Serverless platforms based on Confidential Virtual Machines (CVMs) have been recently proposed to address the privacy problems with serverless functions, while achieving low latency.
Unfortunately, our study indicates that to achieve these properties, existing proposals impose non-trivial requirements in terms of infrastructure changes and platform memory.
\sys is an alternate serverless platform design that does not require infrastructure changes and significantly reduces platform memory requirements.
The platform is designed using two key components: 
(1) a {function isolation framework inside a CVM based on kernel deprivileging features} that minimize infrastructure requirements,
and (2) a layer-wise caching methodology and algorithm that effectively uses a small in-memory function cache.
Our evaluation indicates that \sys can significantly reduce both platform requirements for deployment and function memory consumption.

\end{abstract}

\section{Introduction}
\label{s:intro}

Serverless computing streamlines the deployment of highly-distributed applications in cloud machines~(e.g., data analytics and inference~\cite{faastjs,corral,pocket,lambda,shuffling,serverless-ibm,serverless-ml2,cirrus}),
but also poses unprecedented privacy risks. 
In serverless computing platforms, developers provide only the code and data for individual units of application logic, called {\em serverless functions}. 
The cloud provider automatically manages other aspects of function execution, like memory management, compilation, and dependency management.
Unfortunately, since serverless functions are entirely managed by providers, the control over sensitive information provided to these functions (e.g., patient data provided for analytics by hospitals)
becomes elusive. 

To regain data privacy in serverless computing, confidential serverless platforms~\cite{google-gke} have been proposed, which leverage {\em hardware-assisted trusted execution environments} to isolate function instances from the remaining cloud machine~(\autoref{s:motivation}).
While different iterations of such platforms have been proposed over the years~\cite{rewind,reusable,sgx-serverless}, the most common current approach is to leverage {\em Confidential Virtual Machines} (CVMs), a CPU hardware technology on cloud machines that isolates guest virtual machines from the host infrastructure~\cite{tdx,arm-cca,sev,sev-snp-whitepaper}.
This is primarily also because CVMs are widely-deployed in all modern clouds~\cite{aws-nitro,azure-cc}.

Recent research~\cite{wallet,cofunc} has made notable strides to make confidential serverless platforms~(based on CVMs) more practical.
In particular, a notorious problem for {\em typically short-lived} serverless functions is the {\em cold start latency}---the time it takes to start executing a serverless function.
This problem is exacerbated in CVMs, because they are significantly slower to start than traditional containers or virtual machines, with the additional slowdown from measuring the loaded guest environment (i.e., creating and extending a SHA-256 hash) required for future attestation by remote function users.

Despite aforementioned strides, our analysis on proposed solutions reveals three significant gaps in terms of practical cloud deployment~(\autoref{opportunities}).
First, relying on monolithic kernels like Linux for function isolation inside CVMs expands the Trusted Computing Base (TCB); thus, prior work relies on minimized {micro-kernel} implementations or security monitors to isolate function instances.
Unfortunately, the proposed approaches {require complex modifications to the pre-existing cloud infrastructure.} 
Second, to achieve efficiency, existing systems require large snapshots of functions (termed {\em Zygotes}) to be kept in memory at all times, which significantly increases the memory requirements of the platform.
For instance, our evaluation on eight well-known industry-standard benchmarks~\cite{copik2021sebs}) shows that merely 21 zygotes require $\simeq$5.15GiB of memory, while a real-world deployment would have to keep tens of thousands of diverse zygotes.
Third, when {\em Zygotes} must be evicted (e.g., high memory pressure situations), proposed caching schemes like Least Recently Used (LRU) result in poor invocation latencies.

This paper presents \sys, a confidential serverless platform that addresses the gaps left by prior work to build a deployment-friendly and memory-latency co-optimized solution.
The platform incorporates two main approaches:

\begin{packeditemize}
    \item Function isolation inside a CVM using a security monitor built on the principles of {\em intra-kernel privilege isolation}~\cite{dautenhahn2015nested}, a philosophically-simple monitor design that can be deployed in a self-contained manner within CVMs.

    \item Partial snapshot retention across different layers of the serverless function stack (e.g., language runtimes), directed by the learned invocation patterns of functions and platform memory resource availability.
\end{packeditemize}

Designing a platform around these approaches posed two main challenges.
First, while intra-kernel monitors have been introduced for CVMs~\cite{erebor}, they lack key features like {\em secure Copy-on-Write fork} that are essential to maintain a low startup latency for serverless functions.
Second, partially caching traditional serverless containers using the file system~\cite{rainbowcake} is known, but it is non-trivial to extend these ideas to in-memory functions.

We address the lack of secure {\em Copy-on-Write} (CoW) fork validation in existing intra-kernel monitors by designing new memory management unit-level checks during the kernel's fork system call and page fault handling~(\autoref{s:design:fork}).
At a high-level, these checks require the kernel to explicitly declare its intentions to the monitor during these operations (e.g., what pages are being allocated to a process and at what mappings).
If these intentions are not declared, or declared incorrectly, the monitor does not allow the process to continue and aborts both the parent and child process.
These intentions are validated at key lifecycle events, including page table mapping updates, copying contents between protected pages, and at the end of the fork system call.

For partial snapshot retention and caching of function-running processes, we leverage \sys's secure CoW fork.
At a high-level, we maintain checkpoints during function execution (e.g., when the function has loaded required language runtime) and fork at these times to maintain partial states~(\autoref{s:design:cache}).
Then, we implement a predictive time-series algorithm that considers factors like function arrival times, memory overheads, and startup latency from different cache levels.
Using this algorithm, snapshots are retained/pre-warmed to reduce latency, or evicted to preserve memory.

We built a \sys prototype for Linux kernels running on Intel TDX-based Confidential VMs (CVMs) using $\simeq$8400 SLoC~(\autoref{impl}).
Our prototype consists of several components, including 
an intra-kernel security monitor implementation (built on open-source codebases~\cite{erebor,cki}),
a partial-caching aware Library Operating System (LibOS) implementation, and
a function cache coordinator that integrates with the OpenWhisk serverless platform~\cite{openwhisk:maturity}.
We will open-source all our software artifacts.

Using our prototype, we evaluated \sys on security, performance and memory use.
In terms of security~(\autoref{security}, we analyzed the properties that ensure correct CoW forks and prevent container sharing from leaking sensitive information.
For evaluation~(\autoref{performance}), we selected the same set of benchmarks leveraged by prior work~\cite{wallet} to evaluate the system on micro-benchmarks and a macro study involving real-world function traces from the Microsoft Azure Serverless Function dataset~\cite{serverless-in-the-wild}.
Our results indicate the following aspects:

\begin{packeditemize}
    \item \sys's function startup latency using secure CoW is comparable to prior work~\cite{wallet,cofunc}, showing that our deployment-friendly monitor maintains high performance.

    \item \sys's partial caching and predictive caching algorithm uses 64\% less memory on average compared to prior work, while retaining 19.3\% of the geometric mean function performance.
    
\end{packeditemize}

\section{System and Deployment Model}
This paper is concerned with a {\em dedicated confidential serverless deployment for enterprise customers in cloud machines}.
In serverless terms, such a customer would be a large organization that invokes hundreds of thousands of functions per-hour. 
Enterprise customers are afforded privileges like customized solutions and direct support.
The rest of this section first explains the client's requirements for such a deployment and then presents real-world example scenarios.

\PP{Enterprise client requirements}
The client wants a secure and high performance solution that is also cost-effective.
Specifically, the client is performing computations on sensitive user data (e.g., healthcare or financial data) provided by their downstream customers~(e.g., hospitals).
They want to protect this data from cloud system administrators and other tenants for regulatory reasons (e.g., HIPAA, CCPA) or others.
In terms of performance, they want to minimize the end-to-end function invocation latency to provide rapid response to their downstream customers.
The client may run functions in many different languages (e.g., Python, Java) and may use many different runtime versions of each language (a common occurrence in confidential serverless functions~\cite{rainbowcake}).
To remain cost-effective, they want to minimize the per-function cost.

\PP{Real-world examples}
In practice, there are many companies that utilize serverless computing on sensitive workloads and fit our model. We provide three examples.
(1) LiveWell, a health application by the Zurich Insurance Group, runs on Amazon serverless infrastructure. LiveWell is responsible for collecting and analyzing the daily health habits and biometrics for over 150,000 daily users~\cite{livewell}. 
(2) 
Booz Allen uses AWS Lambda to accelerate and scale agentic AI frameworks designed to assess and mitigate new malware~\cite{awsBoozAllen2025}. Due to their business with the federal government, Booz Allen has taken measures to ensure security and data privacy with Amazon. 
(3) The Netherlands-based Portbase data analytics company processes sensitive and critical logistics information~\cite{awsPortbase2021}. By using AWS Lambda, Portbase is able to achieve high scalability without needing to maintain massive infrastructure.

\section{Background on Confidential Serverless}
\label{s:motivation}

{Confidential serverless computing} addresses the privacy risks posed by serverless functions using {\em hardware-assisted trusted execution environments} (TEEs).
Among deployed technologies, Confidential Virtual Machines (explained next) offer a promising foundation for confidential serverless platforms.

Modern hardware technologies like Intel TDX~\cite{tdx}, AMD SEV~\cite{sev-snp-whitepaper,sev}, and ARM CCA~\cite{arm-cca} 
enable hardware-based isolation of guest virtual machines running on cloud platforms. Such virtual machines are named Confidential Virtual Machines or CVMs.
Within a CVM, the entire guest software stack, including the operating system and applications, are protected from outside components like the host hypervisor, devices, and co-located virtual machines.
In particular, the hardware enforces isolation by encrypting and integrity-protecting guest memory. 
Once launched, the host (hypervisor and other VMs) and external devices cannot access or modify the runtime state of the guest kernel or applications, ensuring runtime confidentiality and integrity.

A key enabler of CVMs is the ability to support remote attestation and establish trust with remote users. 
The CPU produces a signed attestation report containing measurements of the CVM's initial state (e.g., boot firmware, kernel, and configuration). 
Remote verifiers can use this report to confirm the CVM's authenticity and integrity before provisioning secrets. 
The report may also include CVM custom data, enabling secure channel establishment (with the remote verifier) via authenticated key exchange protocols~\cite{erebor}.

\section{Motivation}

\begin{table}[t]
\centering
\scriptsize
\caption{Current state-of-the-art boot times for CVMs.}
\label{tab:cvm-boot-times}
\setlength{\tabcolsep}{3pt}
\renewcommand{\arraystretch}{1.0}
\resizebox{0.9\columnwidth}{!}{%
\begin{tabular}{@{}lrrrrr@{}}
\toprule
\textbf{System} 
& \textbf{Kata CVM} 
& \textbf{Kata microCVM} 
& \textbf{SEVerifast} 
& \textbf{Gramine-TDX} 
\\
\midrule
\textbf{Boot Time (ms)} 
& 6803ms 
& 334ms 
& 325ms 
& 2977ms \\
\bottomrule
\end{tabular}
}
\end{table}

\subsection{Optimized Isolated Function Execution inside a CVM}
\label{approach}

A straightforward confidential serverless solution is to boot-up a serverless function inside a CVM on each invocation.
However, this solution is both slow and costly in our model. 
In particular, serverless functions are {\em short-lived} (e.g., a few seconds).
Launching a CVM requires loading a fresh guest OS and language runtime before execution of the function, and also integrity measurement (i.e., generating a SHA-256 cryptographic hash) of these components to support attestation.
This results in a huge {\em cold-start latency} that dominates function execution time~(\autoref{tab:cvm-boot-times}), a notorious problem for serverless platforms~\cite{agache2020firecracker}.
Moreover, due to security reasons, CVMs cannot share memory with each other.
This creates a problem because many serverless cost optimizations~\cite{rainbowcake,seuss} proposed by prior work depend on memory sharing-based techniques, which now become infeasible.

To address the aforementioned problems, we propose a {\em dedicated isolated container-based deployment for serverless enterprise customers inside a CVM}.
Specifically, the idea is to allow the cloud provider to run a dedicated instance of their serverless platform~(e.g., Lambda, OpenWhisk) inside the CVM and spin-up containers for different functions requested by the enterprise client.
With this approach, each container can be rapidly deployed on-demand inside the CVM, and the provider's software stack can also implement a wide-range of memory optimizations to maintain cost-effectiveness.
However, the obvious problem with this approach is that the CVM's privileged software stack~(including the OS kernel) is now under the control of the cloud provider and can steal user data.
To address this problem, we can leverage {\em CVM secure container solutions}~(explained in the next heading) to isolate userspace containers from the CVM's privileged software stack.

\begin{figure}[t]
  \footnotesize
  \begin{center}
    \centering
    \includegraphics[width=0.95\linewidth]{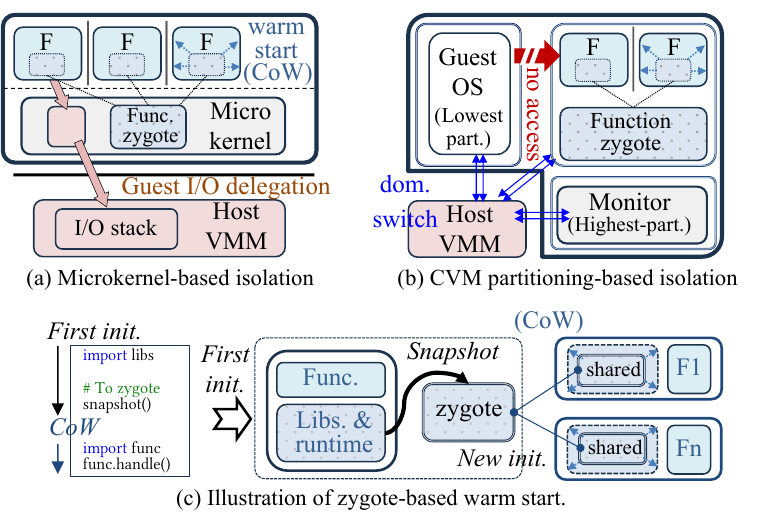}
    \caption {(a) (b) Illustration of existing CVM-based serverless approaches.
    (c) Illustration of the zygote-based warm start mechanism.
    CoW: copy-on-write.
    }
    \label{f:existing-work}
  \end{center}
\end{figure}

\subsection{Potential Existing Solutions and their Limitations}
\label{opportunities}

Prior work instantiates isolated {\em container-like} abstractions inside CVMs using three main design philosophies~(\autoref{f:existing-work}):

\begin{packeditemize}
    \item {\em Trusted micro-kernel}:
    CoFunc~\cite{cofunc} proposes a new {\em micro-kernel} implementation for CVMs that is leveraged to isolate different containers. 
    Specifically, the trusted kernel leverages its control over process contexts (i.e., register states) and page tables to isolate different user containers.

    \item {\em CVM partitioning}:
    Veil~\cite{ahmad2024veil} and Wallet~\cite{wallet} leverage special hardware features in AMD CPUs (i.e., Virtual Machine Privilege Levels or VMPL) to establish a security monitor inside the CVM at higher privileges.
    This monitor isolates different functions from the underlying OS kernel.

    \item {\em Intra-kernel isolation}:
    Erebor~\cite{erebor} leverages the software Nested Kernel~\cite{dautenhahn2015nested} principles to embed a security monitor inside the CVM kernel.
    The {monitor} software gains control over critical system interfaces, including the Memory Management Unit (MMU), while the deprivileged kernel retains responsibility for regular OS services.
    Using its control, the monitor {\em validates security policies on privileged operations} (e.g., checks that the kernel does not map a protected container's memory to other functions).
\end{packeditemize}

Unfortunately, previous solutions present three significant limitations in terms of applicability in our model~(\autoref{t:comparison}).

\begin{table}[t]
  \caption{
  Comparison between \sys (this work) and existing CVM data protection solutions.
  }
  \vspace{-0.2cm}
  \label{t:comparison}
  \setlength\tabcolsep{2.5pt}
  \setlength{\aboverulesep}{0pt}
  \setlength{\belowrulesep}{0pt}
  \begin{adjustbox}{width=\linewidth, center}
\begin{threeparttable}
\begin{tabular}{@{}cccccccccc}
\toprule
\multirow{2}{*}{\textbf{\begin{tabular}[c]{@{}c@{}}Potential \\ Solution\end{tabular}}} 
& \textbf{Isolation} 
& \multicolumn{2}{c}{\textbf{Deployment}} 
& \multicolumn{2}{c}{\textbf{Perform.}} 
& \multicolumn{2}{c}{\textbf{Memory}} \\
\cmidrule(lr){3-4}
\cmidrule(lr){5-6}
\cmidrule(lr){7-8}
& {\bf Supported}
& \textbf{Host} 
& \textbf{HW} 
& \textbf{Feat.\tnote{1}}
& \textbf{Boot} 
& \textbf{Feat.\tnote{2}}
& \textbf{Cons.} 
\\ 
\midrule
Wallet~\cite{wallet}
& Multi-Party
& HV/PV
& VMPL
& F 
& 10.3ms
& -- 
& 2.8$\times$
\\
CoFunc~\cite{cofunc}
& Single-Party 
& HV
& --
& F 
& < 15ms
& -- 
& 2.8$\times$
\\
Erebor~\cite{erebor}
& Multi-Party 
& --
& --
& -- 
& 4.6s
& --
& 2.8$\times$
\\
\rowcolor[gray]{0.85}
\sys 
& Multi-Party 
& --
& --
& F
& 10ms 
& L
& 
1$\times$
\\ 
\bottomrule
\end{tabular}
\begin{tablenotes}
        \item[\textbf{1}]
        {F$\rightarrow$CoW {\em fork} of in-memory Zygotes}
        \item[\textbf{2}]
        {L$\rightarrow$Layered Predictive Caching~(\autoref{s:design:cache})}
\end{tablenotes}
\end{threeparttable}
\end{adjustbox}
\end{table}

\PP{L1: Incompatible isolation or huge deployment changes.}
Our model requires {\em multi-party isolation} between the entity controlling the CVM OS kernel (i.e., cloud provider's serverless platform) and containers.
Unfortunately, this cannot be achieved by solutions like CoFunc~\cite{cofunc} where the OS kernel is the root-of-trust, and can only achieve single-party isolation.

Wallet and CoFunc also require platform-specific mechanisms and host-side modifications, both of which hinder portability and practical deployment. 
To maintain a security monitor, Wallet relies on VMPL to partition the CVM into lower and higher privilege domains.
Using VMPLs in this way not only requires changes to the host hypervisor (e.g., to allow custom switching between different domains) but also the paravisor, a new cloud provider-controlled CVM component that enables CVM migration and TPM services~\cite{erebor}.
This hurts usability and also ties the design to the AMD SEV-SNP platform. 
CoFunc, in contrast, avoids VMPLs but shifts core OS services (e.g., filesystem and network I/O stack) to the host hypervisor.
This is to preserve a minimal in-guest microkernel and maintain a small TCB for intra-CVM isolation.
Such a choice also mandates host environment modifications, complicating not only deployment but migration.

\PP{L2: Large performance slowdown from runtime attestation.}
In our model, the remote users would want to {\em individually attest} isolated functions, but this creates a challenge: the function instance~(including language runtimes and libraries) must be measured, resulting in significant startup delays.
To minimize this delay, Wallet and CoFunc propose the concept of in-memory {\em Zygotes}.
A zygote is the {\bf snapshot} of an initialized and pre-measured function runtime, library dependencies, and even its code state (as illustrated in \autoref{f:existing-work}{\bf c}).
Whenever a function is requested, it can be launched using a fast {\em Copy-on-Write (CoW) fork}~\cite{fork-in-the-road}
\footnote{Like prior work, we use the term \cc{fork} system call family to denote the combined semantics of \cc{fork} + \cc{execv}~\cite{fork-in-the-road}.} 
based on its in-memory zygote, instead of loading and measuring the function from scratch.

Unfortunately, Erebor (and other intra-kernel monitors~\cite{dautenhahn2015nested}) lack the runtime validation required to support {\em fork}.
Please note that we cannot simply import the fork functionality into an intra-kernel monitor (like other monitors do~\cite{wallet}), since this goes against the design philosophy of such monitors---i.e., only carefully verify that the OS kernel performs an operation correctly, instead of re-implementing functionality inside the monitor.
Unfortunately, lacking this support, Erebor incurs a significant function boot latency---i.e., 4.6s on average in our experiments using functions described in~\autoref{performance}, which is orders of magnitude slower than other systems.

\begin{figure}[h!]
  \footnotesize
  \begin{center}
    \centering
  \includegraphics[width=0.95\columnwidth]{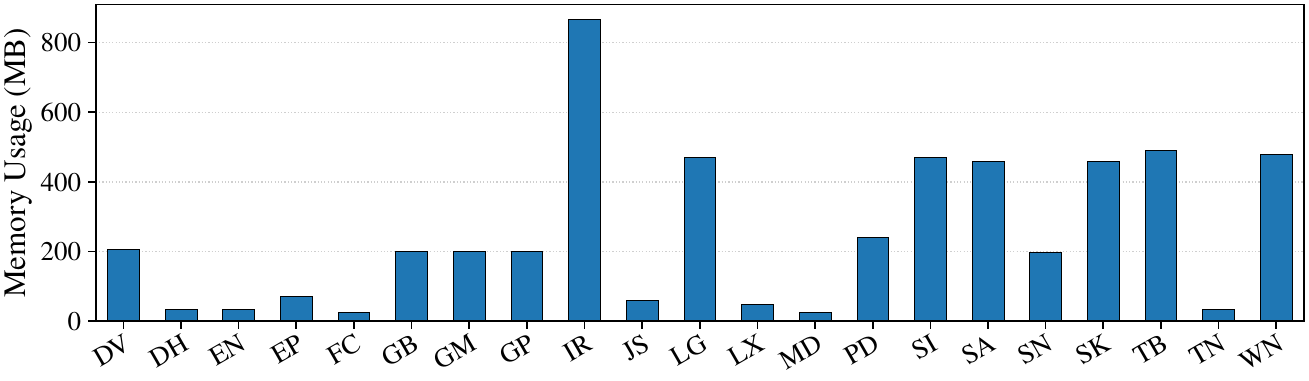}
    \caption {Memory consumed by SeBS Python {\em Zygotes}~\cite{copik2021sebs}.
    }
   \label{fig:zygote}
  \end{center}
\end{figure}

\PP{L3: Prohibitive memory caching and runtime utilization.}
{\em Zygotes} snapshot the initialized state of each function into memory, enabling sharing of common code and data across functions (using CoW), but the snapshots nevertheless still consume considerable memory.
To understand this overhead, we deployed and measured 21 different Python-based functions (from the popular SeBS benchmark~\cite{copik2021sebs} used by Wallet~\cite{wallet}), CoFunc~\cite{cofunc}, and RainbowCake~\cite{rainbowcake}).
\autoref{fig:zygote} reports the observed memory consumption: a single function’s zygote can occupy hundreds of megabytes (totaling $\sim$5.14GiB for only 21 zygotes).
Consider that serverless workloads are highly diverse:
for instance, different Python functions may depend on distinct runtime versions, requiring separate (large) zygotes for each configuration.
To achieve the lowest startup latency, tens of thousands of Zygotes may need to be {\em kept-alive} in memory, raising a non-trivial scalability (and monetary cost) problem for cloud providers.

Inevitably, function {\em Zygotes} will need to be evicted from the in-memory function cache to reduce memory costs due to low function utilization or from high memory pressure in durations of high platform utilization.
This is a persistent reality for traditional serverless platforms running containers or virtual machines~\cite{seuss,rainbowcake,agache2020firecracker}.
Existing solutions do not focus on this problem.
A naive solution would be to leverage a basic scheme like the one implemented in OpenWhisk~\cite{openwhisk:maturity}---loading functions on-demand and keeping them in-memory until a timeout is reached.
In our experiments~(\autoref{performance}), we show that this approach consumes around $2.8\times$ more memory during function execution than our optimized algorithm.

\section{Threat Model}
\label{s:threat-model}

Consistent with the confidential serverless computing model~\cite{wallet,cofunc}, we assume {\em the cloud machine's host OS and hypervisor is untrusted} and may attempt to collect sensitive data provided to the functions.
To do so, the host may exploit the CVM-host interface (e.g., I/O and network stack)~\cite{ghci} to infer data, or intentionally pretend to be a user and run malicious function request workloads to access other (if any co-located) functions and leak data (explained below).
In addition, since our solution~(\autoref{s:design}) is based on a security monitor, we assume that the CVM's kernel is under the control of an adversary (i.e., the cloud provider) and hence untrusted.

In our system model, the enterprise client may be computing on data belonging to different downstream customers in each function.
Therefore, each function must be executed in an isolated environment.
The function developer is trusted by the enterprise client.
Developers supply the function code and a self-contained {\em function manifest} that lists the dependencies required by the function (i.e., language runtimes and libraries) and the integrity measurement of these dependencies (e.g., SHA-256 hashes). 
Since computational acceleration through external components (e.g., GPUs and FPGAs) is an ongoing area of research for Confidential VMs, we assume the provided serverless functions run solely on CPUs.

\PP{Out-of-Scope}
Our work does not consider information leakage through through micro-architectural side channels~\cite{llc-practical,aes-sgx,sge,sgx-branch} or hardware defects~\cite{spectre,meltdown,sgx-spectre,plundervolt}.
For the latter, we assume that the hardware has been patched to the latest version to resolve these issues.
Moreover, we do not address denial-of-service and physical attacks.

\section{\sys Design}
\label{s:design}

\sys is a confidential serverless solution for enterprise customers that addresses identified limitations in terms of {\em platform adaptation} and {\em memory-latency co-optimization}~(\autoref{opportunities}).

Like prior work, \sys instantiates a security monitor based on {\em intra-kernel isolation} to protect the client's functions from the CVM software stack.
We chose this design since it neither requires CVM partitioning features (e.g., VMPL) nor requires delegation of the I/O stack from the host hypervisor~(addressing {\bf L1}).
Our implementation of the monitor is based on recent research that has adapted Nested Kernel principles for CVMs~\cite{erebor,cki}.
These systems allow the creation of {\em Trusted Processes} (akin to SGX enclaves~\cite{sgx-hasp}) that are isolated from  the CVM software (including kernel).

\sys deploys and isolates each serverless function requested by the client within a {\em Trusted Process}.
Functions are executed in processes containing a {\em Library Operating System} (LibOS) to provide a container-like environment that is common in cloud systems~\cite{gramine-tdx,haven}.
Integrity measurement and attestation follow the LibOS two-stage approach~\cite{graphene-sgx}.
In particular, the monitor measures the LibOS runtime and its manifest file (which specifies the runtime and dependencies), and reports this to the remote user.
The (correctly-loaded) LibOS then uses the manifest to measure and validate runtime components and libraries (loaded through calls to the operating system), before executing the function
and accepting user data.

In the system's design, we make three key contributions: 

\begin{packeditemize}
    \item Strictly following the philosophy of intra-kernel monitors, we propose a set of design extensions~(\autoref{s:design:fork}) 
    to validate the correctness of secure Copy-on-Write~(CoW) fork of a {\em Trusted Process}, which is essential for efficient function caching and fast startup~(addressing {\bf L2}).

    \item To optimize cold-start latency while reducing the number of function snapshots~({\em Zygote}) kept in memory~(addressing {\bf L3b}), we propose a partial caching approach that selectively retains, evicts, or pre-warms a {\em Zygote} based on invocation patterns and resource availability (\WC{1}-\WC{2}).
\end{packeditemize}

\subsection{Intra-Kernel Monitor Extensions for Fork Validation}
\label{s:design:fork}

This section explains how we extend Linux intra-kernel monitor designs~\cite{dautenhahn2015nested,erebor,cki} to support secure Copy-on-Write~(CoW) fork of a {\em Trusted Process}, which is essential for efficient function caching and fast startup. 
To support these operations, \sys implements new security monitor calls (SMCs)~(\autoref{t:fork-apis}), which is the interface by which the kernel calls the monitor for privileged operations, and instruments the kernel during fork syscalls and page faults to uphold security invariants. 

At a high level, \sysmonitor protects the memory regions of each trusted process (as well as its forks) during and after fork, preventing any access from untrusted components such as the OS kernel.
That being said, a trusted process's memory address space is composed solely of protected physical pages.
These pages can be accessible only to processes authorized by \sysmonitor (explained below).

\begin{table}[t]
\centering
\caption{
Sensitive privileged instructions and descriptions.
$\bigtriangleup$ denotes pre-existing SMCs updated with new invariants.
}
\vspace{-0.1cm}
\label{t:fork-apis}
\setlength{\tabcolsep}{3pt}
\setlength{\aboverulesep}{0.05pt}
\setlength{\belowrulesep}{0.05pt}
\renewcommand{\arraystretch}{1.0}
\begin{adjustbox}{width=\linewidth}
\begin{tabular}{@{}l|cl@{}}
\toprule
\textbf{SMC API} & \textbf{SMC usage description} \\ 
\midrule
\begin{tabular}[c]{@{}c@{}} \cc{initialize\_fork} \end{tabular} & \begin{tabular}[c]{@{}l@{}} Signals the intent to fork a protected process.\end{tabular} 
\\ 
\begin{tabular}[c]{@{}c@{}} \cc{declare\_page}~($\bigtriangleup$) \end{tabular} & \begin{tabular}[c]{@{}l@{}} Declare a new page table or trusted process page. \end{tabular} 
\\
\begin{tabular}[c]{@{}c@{}} \cc{update\_mapping}~($\bigtriangleup$) \end{tabular} & \begin{tabular}[c]{@{}l@{}} Update a (L1-L4) page table entry's current mapping\end{tabular} 
\\ 
\begin{tabular}[c]{@{}c@{}} \cc{protected\_copy} \end{tabular} & \begin{tabular}[c]{@{}l@{}} Copy from an existing page to a newly-declared page \end{tabular} 
\\ 
\begin{tabular}[c]{@{}c@{}} \cc{declare\_cow\_page} \end{tabular} & \begin{tabular}[c]{@{}l@{}} Mark a protected physical page as Copy-on-Write.\end{tabular} 
\\ 
\begin{tabular}[c]{@{}c@{}} \cc{finalize\_fork} \end{tabular} & \begin{tabular}[c]{@{}l@{}} Signals the end of the fork system call operations.\end{tabular} 
\\ 
\bottomrule
\end{tabular}
\end{adjustbox}
\vspace{-0.2cm}
\end{table}

\PP{Instrumentation during {\em fork} system call}
The initial \cc{fork} call is used by the kernel to set up the child processes' page tables and mark physical pages to be {\em copy-on-write}.
At a high-level, we would need to uphold the following security invariants ({\em Inv1 - Inv4}) during these operations:

\begin{packeditemize}
    \item 
    \vspace{0.05em}
    \noindent
    {\em Inv1:} 
    Protected physical pages can only be shared between processes that maintain an explicit family relationship.
    
    \item 
    \vspace{0.05em}
    \noindent
    {\em Inv2:}
    Protected physical pages can only be mapped at the
    same virtual addresses across all related processes.
    
    \item 
    \vspace{0.05em}
    \noindent
    {\em Inv3:}
    Shared protected physical pages that are to be mapped Copy-on-Write (CoW) must be explicitly declared.

    \item
    \vspace{0.05em}
    \noindent
    {\em Inv4:}
    For a fork, all its related processes must have an identical number of protected physical pages at the end of that fork call.
\end{packeditemize}

\begin{figure}[!t]
 \centering
 \includegraphics[width=.95\columnwidth]{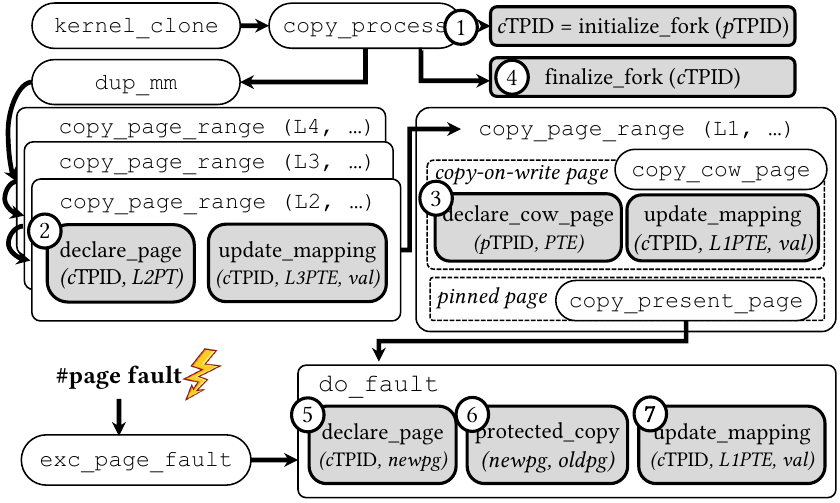}
 \caption{Callgraph of \sys's instrumentation of Linux's process creation and pagefault pathway to validate CoW fork.}
 \label{f:fork-callgraph}
\end{figure}

\vspace{0.15em}
\autoref{f:fork-callgraph}~(\WC{1}--\WC{4}) illustrates how we instrument the kernel's \cc{fork} callgraph to help \sys track important events and validate these invariants (explained in the next heading).

Initially, the kernel executes a secure monitor call (\cc{fork\_initialize}) to specify that a fork operation from a parent is expected.
Without this call, no fork-related operation is allowed by the monitor on any protected pages belonging to the parent process~(\WC{1}).
At this call, the monitor creates an internal enclave structure, and specifies a relationship between the parent and to-be-created child process.
The call returns a new child process identifier that is linked to the process' page table base address (\cc{CR3}).

The next step of the process is to sequentially populate the newly-created child's page tables, starting from the root page table to the lowest level~(\WC{2}).
Following the Nested Kernel principles~\cite{dautenhahn2015nested}, all newly-created page tables are declared by the kernel with the monitor~(using \cc{declare\_page}).
This is to ensure page table entries are protected (i.e., write-restricted) from modification by the kernel.
The monitor validates that all page tables are declared when a modification is requested.
For instance, an entry belonging to the L4 page table is only modified by the kernel to point to a physical page that was declared to be an L3 page table, and so on.

The handling of pages at the lowest-level of page tables is different from the upper levels~(\WC{3}).
At this stage, the kernel marks all {\em non-pinned} pages as CoW in the parent and child.
We instrument the kernel to explicitly declare which physical page is CoW-marked (using \cc{declare\_cow\_page}).
Only pages that belong to processes with family relationships are allowed to be {\em double-mapped} to both page table entries of the parent and child process.
This declaration is also validated when the  kernel requests the {\em read-only} mapping of the CoW page into the child's page tables.
At this point, the monitor also checks the virtual address for consistency~(i.e., {\em Inv2}).
Note that the parent must also request a change to the parent's page tables to ensure {\em read-only} protection, otherwise the fork termination process (explained below) fails.

Once the fork system call process is completed, the kernel must explicitly call \cc{finalize\_fork}~(\WC{4}), otherwise the monitor never schedules the newly-created child process.
At the finalization step, the monitor runs the following checks to ensure invariants are upheld.
First, it scans the page tables to check that all CoW-marked physical pages are {\em read-only} in both the parent and child page tables.
Second, it checks that the parent and child processes have the same number of mapped protected physical pages.
As explained above, we already ensure that each page is mapped at consistent addresses. 
Hence, we can trust that the newly-created child is a {\em correct} duplicate of the parent process' page tables.

\PP{Instrumentation during page faults}
The kernel maps new pages into the child process both during the initial fork call or during page faults.
During these operations, \sys enforces the following {\em additional} security invariants:

\begin{packeditemize}
    \item 
    {\em Inv5:} Contents from protected physical pages can only be copied into other protected physical pages.

    \item 
    {\em Inv6:} After copy, the new protected physical page can only be mapped to the virtual address of the original page.
\end{packeditemize}

\vspace{0.15em}
\autoref{f:fork-callgraph}~(\WC{5}--\WC{7}) illustrates how the kernel's page fault handling pathway is instrumented to uphold these invariants.
In particular, when a page fault is triggered (either by the hardware or software), the kernel must first declare a new protected physical page for the parent or child process~(\WC{5}).
The monitor checks that the page is valid (i.e., was not previously declared as a protected physical page and is not being used elsewhere in page tables).
If validation checks pass, the page is marked as belonging to the process.

By default, based on intra-kernel monitor principles, the kernel cannot directly access a protected physical page.
Therefore, the kernel must execute an SMC~(\WC{6}) (namely \cc{protected\_copy}) to request copy of contents between two protected pages.
At this call, the monitor checks the ownership of each page, and only initiates copy between pages that belong to processes related to each other.
Moreover, the monitor internally updates the page attributes of the new page with the virtual address of the old one.
Finally, the kernel will request an update to the process mappings (in page tables) to reflect the new changes~(\WC{7}).
At this point, \sys implements several validation checks to uphold all invariants.
These checks are described in the next heading.

\begin{figure}[!t]
 \centering
 \includegraphics[width=.95\columnwidth]{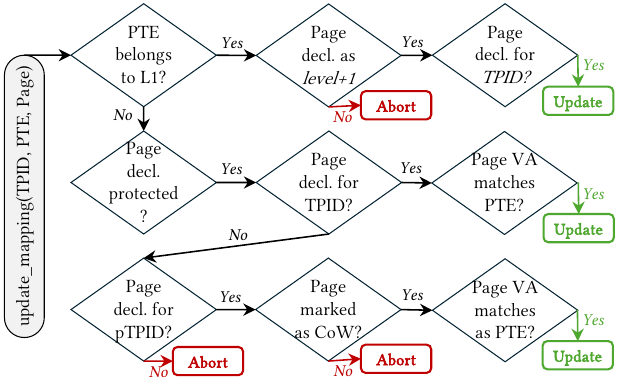}
 \caption{Simplified Finite State Machine of decisions made during \cc{update\_mapping} SMC by \sysmonitor.}
 \label{f:page-validation}
\end{figure}

\PP{Validation finite state machine for mappings}
\sys leverages the instrumented events to validate mappings of the child and parent process during \cc{update\_mapping} and \cc{finalize\_fork} SMCs.
The checks for the latter have already been explained previously in this section.
\autoref{f:page-validation} illustrates the checks made by \sysmonitor during \cc{update\_mapping} as a finite-state machine.
At a high-level, the monitor requires all page table pages to be declared and maps higher-level (i.e., L4--L2) page tables only to physical pages that are declared to lower-level page tables and the same process identifier.
For lower-level (L1) page table entries that map to protected physical pages declared for the current process and the correct virtual address (if pre-assigned during \cc{protected\_copy}), the mapping is permitted.
If the physical page is declared for a different identifier and (a) the process is not declared as a fork or (b) the page is not declared as CoW, the mapping is aborted.
If the page is declared CoW, but the parent identifier is different, the mapping is also aborted.

\subsection{Layered Predictive Function Process Caching}
\label{s:design:cache}

Inspired by recent work on partial container caching using the virtual file system~\cite{rainbowcake,seuss}, we propose to partially cache a pre-warmed trusted process based on the invocation patterns of incoming functions and system configurations.
Unlike prior work, our caching approach does not rely on the file system, but rather a trusted library OS and secure forks.
This section rationalizes \sys's cache layers and describes the algorithm leveraged to maintain contents of these caches.

\PP{Function caching layers and rationale}
\sys maintains function instances in three layers of caches: (1) the bare layer, (2) the (language) runtime layer, and (3) the warm layer. 
During execution, depending on function invocation patterns and cache state, the platform will fork an instance from one of these layers.
We describe the layers below:

\begin{packeditemize}
\item {\em Bare {\bf (B)} layer.} 
This cache layer contains forked function instances that have each loaded the Library OS (LibOS) and standard GNU C libraries in a {\em Trusted Process}.
In particular, a {\em Bare} layer instance is initialized from a secure fork made from a {\em Parent} LibOS instance.
The parent instance is always kept in-memory to permit fast forks.
Bare instances consume a very small amount of memory, and each instance can be {\em warmed-up to run any serverless function requested by users}.

\item {\em Runtime {\bf (R)} layer.} 
This cache layer contains bare LibOS instances that have each loaded a language runtime (e.g., Python or Java interpreter), its dependencies, and standard libraries into memory. 
Note that even within the same language, different versions of the runtime may be used by different functions.
For instance, AWS allows 4 versions of Python-based serverless functions~\cite{aws-python}.
As such, {\em we consider each version a different runtime}.
An instance within this layer can warm up and run any function written for its pre-warmed language runtime, thus permitting faster booting of functions written for such runtimes.
The size of each instance in this layer will be variable, depending on what language runtimes are maintained.

\item {\em Warm {\bf (W)} layer.} The function instances in this layer have already loaded all the required dependencies to run {\em one} specific user-provided function.
As such, this layer is identical to the {\em Zygotes} proposed by prior confidential serverless systems~\cite{wallet,cofunc}.
Each instance can load a user's provided data and directly execute the function, resulting in the lowest latency of the invocation.

\end{packeditemize}

\begin{figure}[!t]
 \centering
 \includegraphics[width=.95\columnwidth]{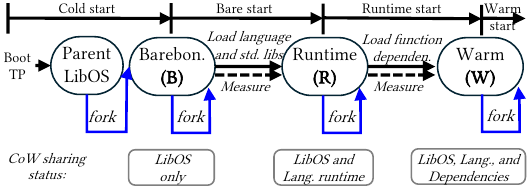}
 \caption{\sys's partial function caching layers, including cost to start from each layer and allowed CoW sharing.}
 \label{f:cache-states}
\end{figure}

\vspace{0.1em}
\autoref{f:cache-states} illustrates how function instances in each of these layers are initialized, including the (high-level) cost associated with serving a function request from different layers.
Loading from the {\em Warm} layer incurs the smallest delay, while lower caching layers incur sequentially more overheads due to additional process creation, library loading, and measurements.
An important caveat to note is the Copy-on-Write (CoW) sharing for each layer that has implications for memory usage.
In particular, depending on the layer, only certain function components can be shared in CoW manner.
For instance, functions forked from the {\em Bare} can only share the LibOS in CoW manner.
Please refer to~\autoref{performance} for a more detailed quantitative breakdown of these differences.

\PP{Predictive cache warmup and eviction}
To reduce memory waste while minimizing startup latency, we rely on a predictive {\em time-series algorithm} for function invocation patterns~\cite{rainbowcake}.
The higher-order goal is to analyze sequential function invocation patterns over time and use them to predict when specific functions may be invoked. 
For predictions, each function is modeled as a Poisson distribution, a common approach to describe arrival patterns~\cite{dudin2020theory,gardner2015reducing,joshi2014delay}.
Based on these predictions, \syshelper keeps the function in {\em Warm} state beforehand, while gradually evicting instances and relying on lower memory-efficient layers during times when certain functions are not being requested by users.

Whenever a function is requested, the coordinator forks a new instance from the cache layer that would incur the least startup latency (i.e., preferring {\em Warm} $\rightarrow$ {\em Runtime} $\rightarrow$ {\em Bare}).
If the function is forked from a layer other than {\em Warm}, the function instance is programmed to also emit a fork as soon as a layer is warmed-up.
For instance, as {\em Bare} is warmed up to {\em Runtime}, a fork is executed and similar at the {\em Warm} state.
This allows the system to maintain pre-warmed function and language instances for future requests.
In principle, \sys only needs {\em one} {\em Bare} instance and {\em one} {\em Runtime} instance (per-language runtime).
However, this results in scenarios where functions are {\em sequentially} waiting for forks to complete.
Thus, the coordinator spins-up new {\em Bare} and {\em Runtime} instances.

The function instances within these layers are kept-alive until the {\em Time-To-Live} (TTL) calculation that is based on two factors: (1) the {\em Inter-Arrival Time} (IAT) that the Poisson distribution predicts and (2) a platform provider-configured cost function that takes into account the function startup latency and memory footprint.
Even when no requests are currently arriving, the coordinator prepares for future requests using the {\em Inter-Arrival Time} (IAT) by warming up {\em Runtime} and {\em Warm} layers.

\section{Implementation}
\label{impl}

This section describes how we built a prototype of \sys for the Linux kernel and Intel TDX-based CVMs~\cite{tdx}.

\PP{\sysmonitor}
Our monitor implementation is built by extending Erebor~\cite{erebor}.
We chose this implementation because of Erebor implements all the pre-existing Nested Kernel principles, and provides the ability to create trusted processes for our serverless functions.
We extended the monitor's implementation with 
1210 lines of code to support the 6 new SMCs that enable secure Copy-on-Write fork~(\autoref{t:fork-apis}).
We rely on Erebor’s existing secure context switches between monitor and the kernel to maintain security for our SMCs.

\PP{Kernel instrumentation}
We instrumented the kernel's source code to call newly-designed SMCs that register fork operations and validates them~(\autoref{s:design:fork}) using around 500 lines of code changes.
Our prototype requires the host machine and CVM to run a kernel compatible with Intel PKS~\cite{pks} and TDX, and thus we used the Intel-provided kernel version 6.8.0.

\PP{Function Library OS~(LibOS)}
For the Library OS on which our functions execute, we leverage the industry-standard Gramine~\cite{gramine-tdx,graphene} like the recent confidential serverless system~\cite{wallet}.
The Gramine LibOS provides a high degree of POSIX compatibility, and thus the ability to run different functions.
We modified the base LibOS implementation to support forking at different layers of the function~(\autoref{s:design:cache}).
In total, we added 1205 lines to the Gramine LibOS to support the aforementioned operations.

\PP{\syshelper}
The coordinator is implemented in 1700 Lines of Code (LoC) of Python. We use a flash server setup to handle invocation requests made by the RainbowCake OpenWhisk orchestrator~(explained next). 
During boot, the coordinator starts the Parent LibOS and the coordinates with it to create new function instances using the fork syscall. 
As new instances get created, the function layer caches are populated~(using forked processes) and the coordinator precisely tracks which process is at what layer.
When spawning a new function instance to handle an invocation, the coordinator iteratively checks all cache layers based on our algorithm~(\autoref{s:design:cache}). 

\PP{Function orchestrator}
To invoke function instances for our evaluation, we use the OpenWhisk-based RainbowCake orchestrator~\cite{openwhisk:maturity,rainbowcake}. 
The orchestrator is written in Scala code. 
To implement our containerization system, we implemented around 1200 lines of Scala code. 
The purpose of the added code is to allow OpenWhisk a way to communicate with the VM server. Openwisk requires that a container-like system be used with it, so we implemented a Scala shim layer that passes commands directly to our VM server, while preserving the internal data structures required by OpenWhisk.

\section{Security Analysis}
\label{security}
This section analyzes key security features of the \sys platform and reasons about their ability to prevent attacks.

\PP{Validating the correctness of Copy-on-Write {fork}}
The attacker's goal is to either break the integrity of the parent and child process' page table mappings (e.g., insert malicious mappings) or modify page contents during fork system call and subsequent page faults.
\sysmonitor prevents these attacks through invariants~({\em Inv1--Inv6}) described in~\autoref{s:design:fork}.

\PP{Claim 1} {\em The attacker cannot insert malicious translations into the parent or child functions.}

When forking a new process, the kernel must explicitly mark the new (empty) process as related to the parent~({\em Inv1}).
For the child process, the kernel must declare page tables at all levels, and request all new mappings through the monitor. 
The monitor only inserts shared mappings between related processes at the same virtual addresses~({\em Inv2}) and with non-write permissions~({\em Inv3}).
Recall that it does so by internally tracking the attributes of each page using the Nested Kernel principles.
Moreover, at the end of the fork operation, the monitor ensures that both parents and child have the same number of mapped pages, thus ensuring identical state~({\em Inv4}).

\PP{Claim 2} {\em The attacker cannot access data stored in protected pages belonging to the parent or child.}

The deprivileged kernel is restricted by the monitor from directly accessing a protected page~\cite{dautenhahn2015nested}.
As such, the kernel can only request the monitor to copy contents belonging to a protected page. 
At such requests, the monitor enforces that only pages marked as Copy-on-Write (and thus read-only) are allowed to be copied, and only to other pre-declared protected pages belonging to related processes~({\em Inv5}-{\em Inv6}).

\PP{Enabling confidentiality with partial container caching}
The system allows partial container states (namely {\em Bare} and {\em Runtime}) to be used to load any function or any function related to the language runtime, respectively. 
The attacker's goal would be to try and steal user functions or data from such sharing of partially-warmed states.

\PP{Claim 3} {\em Malicious users cannot steal data belonging to other users from cached function instances.}

The partially warmed states are created before any user data is processed, ensuring that the shared {\em Bare} and {\em Runtime} states contain only the LibOS and language runtime, respectively, and no function- or user-specific information.
Each user invocation operates in a freshly forked trusted process derived from these pre-warmed states. 
Once a function is warmed to the {\em User} state and begins handling user inputs, the monitor prohibits any further fork operations from that process, preventing an attacker from cloning a data-bearing function instance.
Moreover, all writable pages for user-level execution are copy-on-write (CoW) from the shared states, ensuring strict separation between functions. 

\PP{Claim 4} {\em The shared states cannot be tampered with by the untrusted OS to leak user data.}

\sys ensures the integrity of all shared states ({\em Bare}, {\em Runtime}, and {\em Warm}) through hash-based measurement, monitor verification, and memory isolation.
During the generation of shared states, the LibOS verifies all loaded components against developer-provided hashes. 
Any attempt by the untrusted OS to substitute a layer at creation (e.g., a malicious Bare or Runtime image) causes hash verification to fail, and the LibOS to abort.
Once cached layers reside in protected memory, the OS cannot modify or replace them, as it has no write access to such memory regions.
\section{Performance Evaluation}
\label{performance}

\subsection{Experimental Setup}

\PP{Specifications.}
We ran all experiments on an Intel® Xeon® 6510P server running Ubuntu 24.04 (Linux v6.8.0), with 32 physical (2.3 GHz) CPU cores, 128GiB DDR5 memory, and 512GB Solid State Drive storage. 
We assigned 8 vCPU cores, 24GiB memory, and 100GB virtualized (virtio) storage disk to a guest virtual machine, also running Ubuntu 24.04.
For CVM support, we used Intel-TDX module version 1.5.05.46 for our with Linux version 6.6 running within the CVM.

\PP{Workloads}
For performance testing, we used the same workloads leveraged by recent confidential serverless work~\cite{wallet}.
In particular, these workloads include a set of Python benchmarks from the Serverless Benchmarking Suite (SeBS)~\cite{copik2021sebs}.
This is a widely-used automated testing suite that has been leveraged for evaluation by many works~\cite{rainbowcake,cofunc}.
Additionally, for a large-scale simulation, we leveraged the Microsoft Azure Functions production traces~\cite{serverless-in-the-wild}.

\PP{Variants.}
All of the following configurations are deployed using a TDX CVM. (1) {\bf \sys} refers to our system, which features partial caching as the main optimization. (2) {\bf Baseline-OD} refers to a guest CVM with access to the aforementioned confidential forking mechanism, which is available for loaded enclaves to use to accelerate invocations. This is a similar system to what was demonstrated in works such as Wallet and CoFunc. {\bf Baseline-OD} does not bring functions into memory until they are invoked. And finally, (3) {\bf Baseline-FC} refers to the same system configuration as {\bf Baseline-OD}, except the cache always contains all of the benchmarked functions. This serves as a {\em best-case} oracle caching mechanism which is always aware of what functions will arrive and when. 

\subsection{Micro-Benchmarks}

\PP{Secure {Copy-on-Write} fork cost}
We evaluated \sys's performance overhead on system events using LMBench~\cite{larry1996lmbench}, a widely-used benchmark suite for low-level Linux system events. 
Specifically, we used LMBench's {\em fork+execve} and {\em pagefault} benchmarks. 
Compared to {\bf Native}, we observed 
{3.48$\times$} 
and 
{5.84$\times$} 
overhead, respectively.
The root cause of this overhead are many Secure Monitor Calls (SMCs) executed for paging, as also reported by prior intra-kernel monitors~\cite{dautenhahn2015nested,erebor}.
Each of these SMCs cost around $\sim$1.2k cycles for secure context switch and restore.
In practice, the added latency (while visible in benchmarks) has negligible impact on process invocation and is amortized at runtime.
For instance, for a program of 
{196}MiB 
size, a CoW fork only takes 
$\sim${10}~ms on average.
This is comparable to the CoW fork implementations of existing confidential serverless systems~\cite{cofunc,wallet}.

\PP{Layered caching breakdown}
Based on invocation patterns, \sys maintains functions in partially-cached states (i.e., {\em Bare}, {\em Runtime}, and {\em Warm}) to optimize memory utilization~(\autoref{s:design:cache}).
To understand the memory saved by keeping functions in these states, as well as the cost incurred to warm them up, we ran experiments using our function dataset.

\begin{figure}[t]
  \footnotesize
  \begin{center}
    \centering
   \includegraphics[width=\columnwidth]{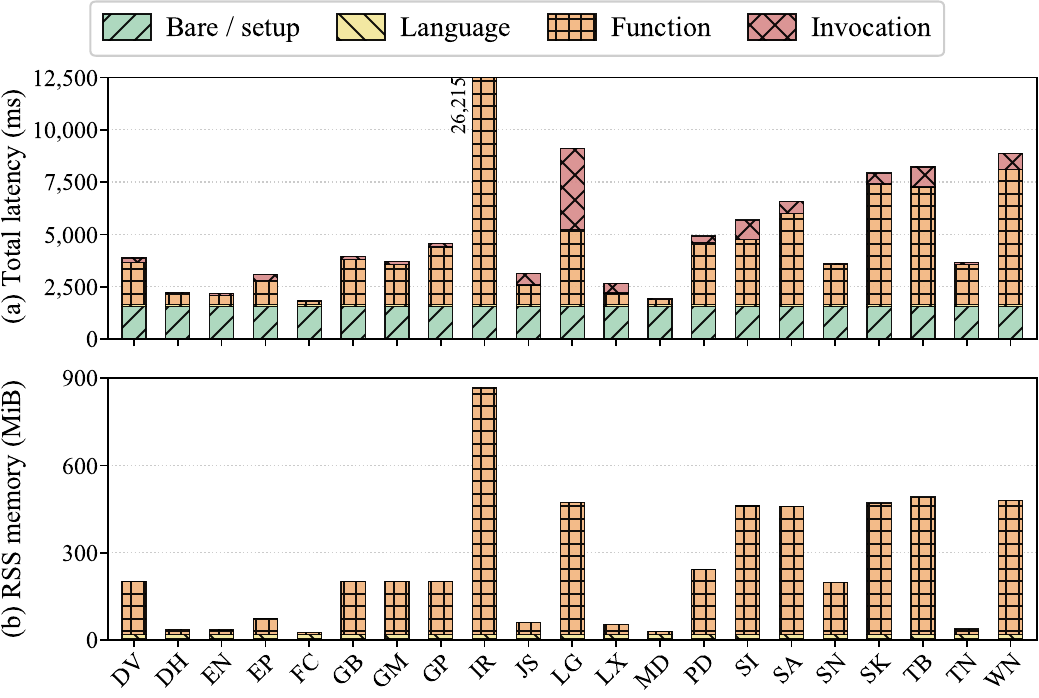}
    \caption {Function (a) invocation latency (b) memory usage in different cache layers.}
   \label{f:cache-memory-latency}
  \end{center}
\end{figure}

\autoref{f:cache-memory-latency}(a) illustrates the warm-up latency across functions.
{\em Bare} and {\em Runtime} states have similar startup costs (on average differing by only $\sim$80ms), as both require initializing core language runtime structures before execution.
At first glance, this might suggest that \sys could maintain only a {\em Bare} instance.
However, {\em Runtime} preserves substantially more copy-on-write (CoW)–shareable memory, as one language runtime can serve many functions, and thus provides much better amortized memory efficiency across workloads.
On the other hand, {\em Warm} significantly reduces latency to tens or hundreds of milliseconds---often 10–20$\times$ faster than Bare and Runtime. 
This improvement comes from restoring a nearly execution-ready process snapshot, which bypasses the expensive dependency-loading path.
Moreover, functions with larger dependency trees (e.g., {\em dna-visualization} or {\em image-recognition}) benefit from the warm state the most, with warm-start latency reductions of up to 25$\times$. 

\autoref{f:cache-memory-latency}(b) further reports the memory footprint of the three states. 
{\em Bare} consumes minimal memory ($\sim$2MB), as it contains only the lightweight LibOS kernel states.
{\em Runtime} requires slightly more memory ($\sim$9MB) to retain the initialized language runtime.
In contrast, {\em Warm} captures application-specific runtime state (program code, initialized stack and heap) and dependency pages, ranging from $25$MB to over $700$MB depending on application complexity. 

\begin{figure}[t]
  \footnotesize
  \begin{center}
    \centering
   \includegraphics[width=0.85\columnwidth]{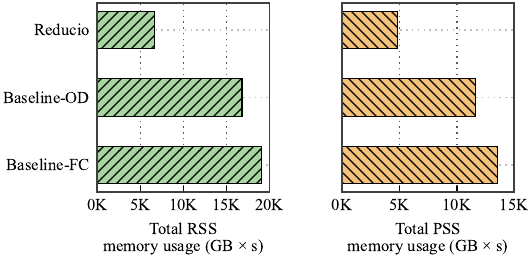}
    \caption {RSS/PSS Memory Usage for each system configuration.}
   \label{f:total-memory-use}
  \end{center}
\end{figure}

\begin{figure}[t]
  \footnotesize
  \begin{center}
    \centering
   \includegraphics[width=\columnwidth]{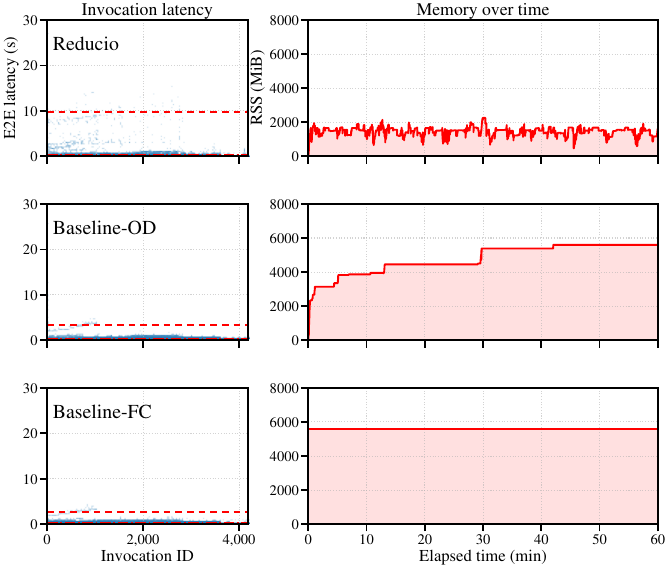}
    \caption {Invocation and Memory over time, depicted here. }
   \label{f:memory-use-latency}
  \end{center}
\end{figure}

\begin{figure}[t]
  \footnotesize
  \begin{center}
    \centering
   \includegraphics[width=\columnwidth]{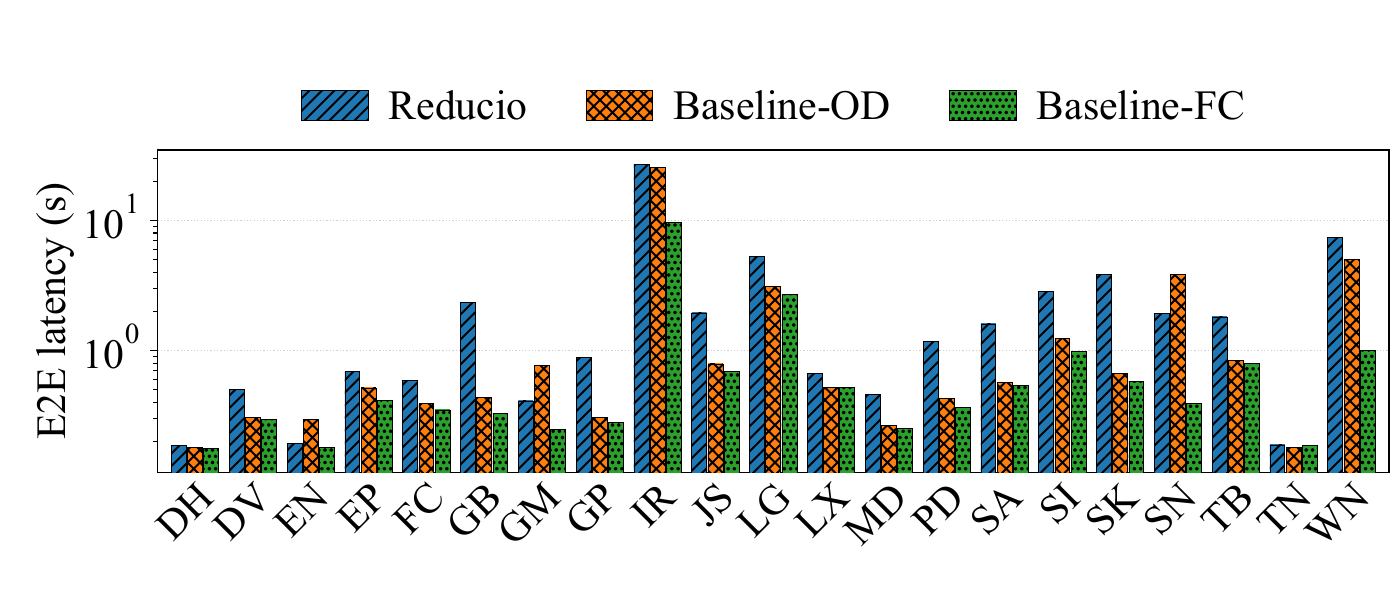}
    \caption {The average end-to-end latencies for each function.}
   \label{f:latency-for-actions}
  \end{center}
\end{figure}

\subsection{Serverless Workloads Analysis} 
For end-to-end evaluation of \sys, we leveraged realistic serverless invocation patterns derived from the Azure Functions traces used in prior work, such as RainbowCake~\cite{rainbowcake}.

\PP{Workload trace setup}
To derive the invocation patterns, we follow a similar methodology to RainbowCake and InstaInfer.
As the functions and their owners are hashed within the Azure dataset, we cannot simply find functions with similar semantics to the functions within the dataset. 
Instead, we associate a function with a trigger type, and find a workload that has the same trigger.   
In order to account for create a diverse set of workloads, we compute the Coefficient of Variance (CoV) across all of the workloads in the data set, and their average invocation rate.
We assigned our benchmarks a CoV and sampled a random span from the Azure dataset for each function.
For our experiments, each benchmark was given a 1 hour invocation trace from the data set that fit the CoV.

\noindent
{\em Invocation Scheduling:} We use an invoker script to schedule the invocation request by taking the minute by minute CSV created by our invocation generator. For a given minute, a function may run 0 to many times. If a function is to be invoked multiple times within a minute, the invoker evenly distributes the invocation rate across that minute. For example, if a function is to be invoke 4 times in that minute, the invoker will call that function once every 15 seconds.

\noindent
{\em Cache Management Parameters:} 
Finally, as the caching mechanism is built on top of RainbowCake's caching management system, we utilize the same parameters for the algorithm as they did. The parameter $\alpha$, which weighs the cost of initialization against memory cost, is set to the RainbowCake optimal, 0.996. The IAT quantile parameter, which measure confidence in container management, is set to 0.8. Lastly, the sliding window parameter set to 6.

\PP{Invocation Trace Latencies}
The left graphs in \autoref{f:memory-use-latency} show
all of the invocations made within their respective experiments. The geometric mean and the p99 latencies have been drawn.
\sys's geometric mean (384 ms) is within 19.9\% of
Baseline-OD (320 ms) and within 19.3\% of Baseline-FC
(322 ms).  REDUCIO will, of course, incur more cold-starts than the single cold-start per function that Baseline-OD has, and the full-cache approach of Baseline-FC.
Meanwhile, \sys’s p99 latency (8385 ms) is considerably large than the p99 latencies of Baseline-OD (3307ms) and Baseline-FC (2687 ms).
\sys will, of course, incur more cold-starts than the single cold-start per function that Baseline-OD has, and the full-cache approach of Baseline-FC. 
As a consequence, the p99 latencies for \sys will always be worse than a system that keeps loaded functions in memory. However, the proximal geometric mean shows that the costs incurred by managing the cache properly can achieve performance that is reasonable compared to an oracle caching mechanism for most invocations.

\PP{Function Latency Breakdown}
\autoref{f:latency-for-actions} plots the average startup and end-to-end latencies for each function. 
Using a function-balanced summary, \sys, Baseline-OD, and Baseline-FC achieve geometric means across functions of the per-function average end-to-end latencies of 1208, 738, and 500 ms, respectively.
The per-function differences are largely attributed the CoV and invocation rate. 
Functions that are invoked in sparse amounts every few minutes perform the worst with predictive caching, while functions that are invoked in bursts benefit the most. 
A good comparison that illustrates this is {\em graph-mst} and {\em graph-bfs}. 
The {\em graph-bfs} function is executed on a trace with a CoV of 1.45, with a medium invocation rate, while {\em graph-mst} is executed with the same rate, but with a CoV of 7.75. 
This means that {\em graph-mst} is infrequently invoked, but when it does, it is invoked in bursts, while {\em graph-mst} is invoked frequently, but with sparse amounts. 
{\em graph-mst} represents a workload that has many invocations per function initialization, while {\em graph-bfs} represents a workload with few invocations per initialization.  

\PP{Memory Usage}
\autoref{f:total-memory-use} shows each system’s memory usage, in terms of gigabyte-seconds. 
\sys has considerably less memory usage. Baseline-OD and Baseline-FC use 2.4x and 2.8x more memory than REDUCIO, respectively. REDUCIO uses 58\% less memory than Baseline-OD and 64\% less memory than Baseline-FC. This shows that predictive caching has a substantial effect on the size of the memory-cache.

\PP{Function Memory Monetary Costs}
The monetary cost of running serverless functions requires two variables: invocation rate and memory size. 
To calculate the cost of memory, cloud providers such as AWS Lambda \cite{aws_lambda} use a step-wise function, dependent on use, which benefits large users.
For functions running on x86 machines, the total cost per GB$\times$S is 0.0000133334\$.
Compared to Benchmark-OD (\$0.16) and Benchmark-FC (\$0.18), running \sys (\$0.06) lowers costs by ~2.4x, and ~2.8x, respectively.

\section{Related Work}
\label{sec:related}

\PP{Privileged software monitoring.}
Designing security monitors that execute at a higher privilege level than the OS is a well-established technique for creating protected process contexts (e.g., enclaves) or enforcing strong kernel isolation. 
One approach is to leverage the hypervisor layer (e.g., VMX) to maintain process-level isolation~\cite{trustvisor,inktag,blackbox,overshadow}, and similar trusted-hypervisor techniques have been used to harden operating systems~\cite{proskurin2020xmp}.
However, in confidential cloud computing, the cloud hypervisor is untrusted, making such designs unsuitable. Instead, \sys aligns with a second line of work based on intra-kernel isolation~\cite{dautenhahn2015nested,erebor,sva,cki,bulkhead}. 
These systems virtualize a higher-privileged monitor within the kernel using hardware memory-protection primitives (e.g., protection keys~\cite{pks}) and privileged-instruction trapping to create a minimal, trusted execution layer.
\sys extends these approaches by supporting secure fork over layered function-caching states.

\PP{Function caching for serverless functions.}
A line of work reduces cold starts by caching function or runtime state and tuning container keep-alive policies.
Shahrad et al. provide a seminal characterization of Azure Functions and show that carefully tuned keep-alive policies for idle containers can substantially reduce cold-start frequency~\cite{serverless-in-the-wild}.
FaaSCache~\cite{faascache} improves on TTL-based keep-alive by adopting a Greedy-Dual–style caching policy, prioritizing which warm containers to retain based on recency, cost, and size.
IceBreaker~\cite{icebreaker} leverages heterogeneous node types and selects where and how to keep functions warm according to invocation probability and node cost.
Another trend is to leverage partial caching, separating the function into different sharable states.
SEUSS~\cite{seuss} deploys functions from unikernel-based snapshots, enabling partial sharing of initialization state across invocations.
RainbowCake~\cite{rainbowcake} applies layer-wise container caching with online keep-alive decisions.
\sys draws inspiration from these caching systems.
It extends layered caching and eviction policies into trusted processes, enabling secure and efficient caching for isolated function processes.

\section{Conclusion}

\sys is a confidential serverless platform that is deployment-friendly and designed from the ground-up to optimize the memory-latency trade-off in existing solutions.
To achieve these properties, the platform combines an intra-kernel monitor design and an intelligent predictive caching approach.

\balance
\bibliographystyle{plain}
\bibliography{p,sslab,conf}
\end{document}